\documentclass[intlimits,twoside,a4paper]{article}

\usepackage{graphicx,epstopdf}

\usepackage[cp1251]{inputenc}

\usepackage[eqsecnum]{cmpj3}

\issue{2017}{20}{2}{23704}
\doinumber{10.5488/CMP.20.23704}

\title[Light absorption coefficient of an ordered array of spherical quantum dot chains]%
{Light absorption coefficient of an ordered array of spherical quantum dot chains}%

\author[V.I. Boichuk, I.V. Bilynskyi, R.I. Pazyuk]{V.I. Boichuk, I.V. Bilynskyi, R.I. Pazyuk}
\address{
Department of Theoretical and Applied Physics \& Computer Simulation,
Ivan Franko Drohobych Pedagogical University,
24 Ivan Franko St., 82100 Drohobych, Ukraine
}

\date{Received March 17, 2017, in final form May 5, 2017}

\begin{document}

\maketitle

\begin{abstract}
We considered intersubband electron transitions in an array of one-dimensional chains of spherical quantum dots in the GaAs/Al$_{x}$Ga$_{1-x}$As semiconductor system. The absorption coefficient caused by these transitions was calculated depending on frequency and polarization of incident light and on Fermi level position, and temperature. We established the existence of two maxima of the absorption coefficient at the edges of the absorption band. It is shown that the absorption coefficient reaches its maximal value at the center of the region between  the $s$-, $p$-like subbands and slightly varies with temperature. The change of the direction of the linearly polarized light wave incident on the chains from perpendicular to parallel leads to a sharp narrowing of the absorption band. It is obtained that the absorption bandwidth increases with the reduction of the quantum dot radius. We also analyzed the dependence of the absorption coefficient of GaAs/Al$_{x}$Ga$_{1-x}$As superlattice on concentration of aluminium in the matrix.
\keywords quantum dot, superlattice, electronic states, density of states, absorption coefficient
\pacs 73.21.Cd, 78.67.Pt
\end{abstract}

\section{Introduction}

Investigations of one-dimensional superlattices (spatially ordered chains) of self-assembled semiconductor quantum dots (QDs) are interesting and potentially practically important in terms of developing the fabrication techniques and theory of physical processes of nanosize heterosystems \cite{Led98, Bim98}.

Semiconductor quantum dots are local areas of semiconductor material about 2--10~nm in size embedded in a wider energy gap semiconductor matrix. Due to the existence of a potential well, charge carriers are confined inside the dots. Due to their small size, quantum dots represent the ultimate case of quantum confinement when the movement of carriers is limited in all three directions. A three-dimensional confinement of carriers leads to a strong spectrum modification of electronic states, so that the electronic structure of a quantum dot brings an analogy with a single atom.

Modification of electron density of states due to three-dimensional confinement effect makes arrays of quantum dots extremely promising for application in a variety of optoelectronic devices \cite{Ara82}. In particular, quantum dot arrays used as an active region of injection lasers are capable of significantly reducing the threshold current density \cite{Dep08}, attaining high temperature stability of the properties \cite{Fath04}, expanding the spectral range achievable in light-emitting devices synthesized on this type of semiconductor substrate~\cite{Max08}. In particular, using arrays of GaAs/AlAs quantum dots made it possible to achieve low threshold lasing on AlAs substrates in the spectral range of the order of $1.3$~{\textmu}m, which opened up prospects of using quantum dot lasers in the systems of optic fiber communication \cite{Zhu08}.

Ordered multiple-layer arrays of vertically coupled QDs were grown by the authors in \cite{Led96, Wan04, Mano04}. The authors of the \cite{Hol07} investigated a superlattice of vertically coupled GaAs QDs in the Al$_{x}$Ga$_{1-x}$As matrix arranged along the elliptic quantum wire. It is shown that the electron energy spectrum in quantum dot superlattices is an alternation of even/odd allowed/forbidden energy minibands. The position and the number of minibands are determined by the quantum dot size. The width of the allowed minibands is defined by the thickness and height of potential barriers.

The absorption coefficient of direct intersubband transitions in a 3D superlattice of a spherical GaAs quantum dot crystal in the matrix Al$_{x}$Ga$_{1-x}$As $(x=0.2, \dots ,1.0)$ is obtained in \cite{Boichuk15}. The calculations show that electronic transitions are allowed between $1s$- and $1p_{z}$-subbands in the dipole approximation. The absorption coefficient is characterized by two peaks attributed to electron transitions with wave vectors in the center and at the edge of the Brillouin zone.

In this paper we investigate the polarization effect of incident light on the absorption coefficient of a one-dimensional array of ordered spherical quantum dot chains. The present calculations based on the method described in \cite{Boichuk15} allow us to obtain the electron energy spectrum of 1$s$- and three 1$p$-subbands for different QD radii and Al concentrations in the Al$_{x}$Ga$_{1-x}$As matrix.

\section{The basic formulae}

\begin{figure}[!h]
\begin{center}
\includegraphics{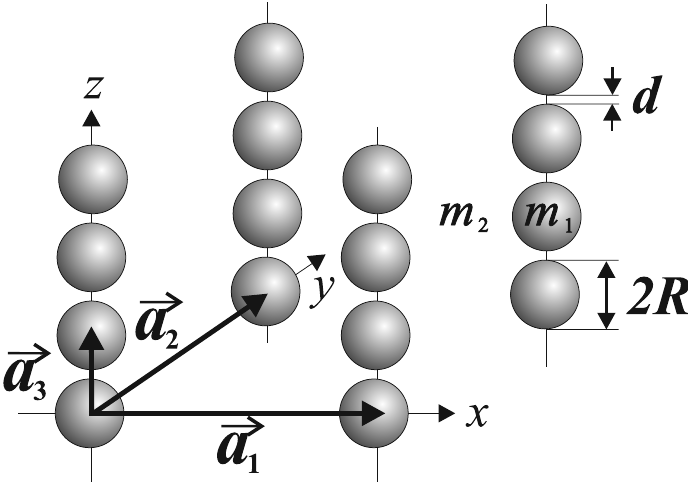}
  \caption{Geometric scheme of one-dimensional superlattices.}\label{Figure1}
\end{center}
\end{figure}

We consider a system of chains of spherical quantum dots arranged along the $Oz$-direction (see figure~\ref{Figure1}).

We assume that distances $d_x$, $d_z$ between the QD chains are so large that quantum dots of adjacent chains can be considered non-interacting. We refer to this structure as a system of one-dimensional superlattices of semiconductor spherical quantum dots. The unit cell of a superlattice has the form of an interval equal to $\left|\vec{a}_{3} \right|\equiv a$, i.e., the distance between QD centers in the direction of the $Oz$-axis.

We investigate optical properties of the system described above. Let us assume that monochromatic electromagnetic wave is incident on the superlattice structure; its polarization is given by vector $\vec{\xi }$ and is characterized by vector potential

\begin{equation}
\label{a}
\vec{A}=-\frac{\ri c}{\omega } \vec{\xi }A_{0} \re^{\ri\left(\vec{\chi }\vec{r}-\omega t\right)}.
\end{equation}

\begin{figure}[!t]
\begin{center}
\includegraphics{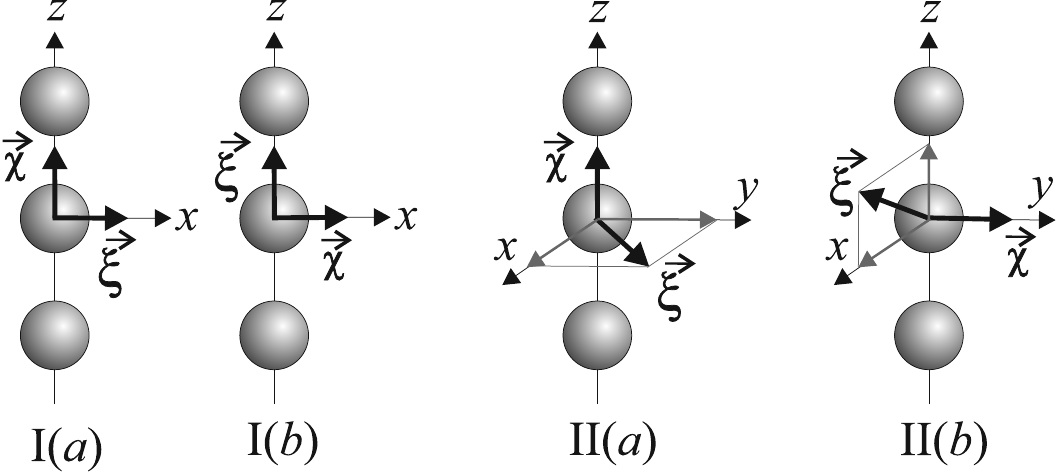}
\caption{Geometry of the experiment: linearly polarized (I) or circularly polarized light (II) is incident (a) parallel to the QD chain; (b) perpendicular to the QD chain.}\label{Figure 2.}
\end{center}
\end{figure}

We consider different possible types of polarization of the incident wave in this system (see figure~\ref{Figure 2.}), focusing only on one chain of the system. In case of linearly polarized light, the polarization vector $\vec{\xi }$ can be directed both perpendicular the QD chain [see figure~\ref{Figure 2.}~I(a)] and along to it [see figure~\ref{Figure 2.}~I(b)]. The wave vector, which gives the direction of electromagnetic wave propagation, is perpendicular to the wave polarization vector ($\vec{\chi }\bot \vec{\xi }$). Similarly, we can consider two directions of incidence of circularly polarized light [see figure~\ref{Figure 2.}~II(a) and~II(b)].

The interaction of electromagnetic wave of a terahertz frequency range with an electronic subsystem of the superlattice leads to electron transitions from a subband $\nu$ of the conduction band $c$ with quantum numbers of the initial state $\lambda =\{c,\nu ,\vec{k}\}$ to the final state $\lambda '=\{c,\nu ',\vec{k}'\}$. The probability of such transitions determines the absorption coefficient that is actually experimentally-observed
\begin{equation}
\label{alfa}
\alpha =\frac{4\left(2\piup e\hbar \right)^{2} N}{Vm_{0}^{2} \omega c\sqrt{\varepsilon } } \sum _{\lambda ,\lambda '}\big|\big(\vec{\xi },\vec{J}_{\lambda \lambda '} \big)\big| ^{2} \left[f\left(\lambda \right)-f\left(\lambda '\right)\right]\delta \left(E_{\lambda '} -E_{\lambda } -\hbar \omega \right),
\end{equation}
where $m_0$ is a free electron mass, $N$ is the number of chains of the length $L$ in the volume $V=S\cdot L$, $J_{\lambda \lambda '} =\int \psi _{\lambda '}^{*} \vec{\nabla }\psi _{\lambda } \rd\vec{r} $ is the matrix element of electron transition from state $\lambda$ to state $\lambda '$ in the dipole approximation, $\psi _{\lambda }^{} =u_{c} \left(\vec{r}\right)\phi _{c\nu \vec{k}} \left(\vec{r}\right)$, $\phi _{c\nu \vec{k}} \left(\vec{r}\right)$ is the envelope wave function of a particle, $u_{c}
\left(\vec{r}\right)$ is the Bloch oscillating function of a particle in the conduction band at the $\Gamma$ point of the Brillouin zone, $\sqrt{\varepsilon } =n$ is the refractive index of the nanocrystal.

If the electron gas in equilibrium is disturbed only by light absorption, distribution functions $f\left(\lambda \right)$ and $f\left(\lambda '\right)$ can be replaced with equilibrium Fermi functions

\begin{equation}
\label{ef}
f_{0} \left(\lambda \right)=\frac{1}{\exp \left(\frac{E_{\lambda } -\mu }{k_{\text B} T} \right)+1} \,,
\end{equation}
$\mu $ is the chemical potential (Fermi level), $E_{\lambda}$ is the energy charge that is determined from the dispersion equation, $k_{\text B} $ is the Boltzmann constant, $T$ is temperature in the system.

When calculating the electron spectrum, we use the effective mass approximation which is valid for not very small QD radii ($R>2$~nm) \cite{Yef85}. Under such conditions, the effective electron masses in a QD ($m_{1}$) and in the matrix ($m_{2}$) are known and equal to those in the corresponding bulk crystals. In the considered heterosystems, lattice parameters and dielectric permittivities of both constituent crystals are very close in values ($a\approx a_{m} $, $\varepsilon \approx \varepsilon _{m} $). Then, the deformation and polarization effects at the interface can be neglected or taken into account by perturbation theory \cite{Boichuk13}.

We investigate the light absorption in case of direct intersubband transitions of electrons and consider the light absorption coefficient of electron transition from the $s$-like to three $p$-like subbands in the dipole approximation $\left(\chi a \ll 1\right)$, where $a$ is the superlattice parameter.

An explicit form of envelope functions in the tight-binding approximation

\begin{equation}
\label{fi}
\phi _{c,\nu ,\vec{k}}^{} \left(\vec{r}\right)=\frac{A(\vec{k})}{\sqrt{a} } \sum _{\vec{n},l,m}C_{l,m}^{\nu } \re^{\ri\vec{k}\vec{n}} \Psi _{l,m}^{0} (\vec{r}-\vec{n}), \qquad \nu =1,2,3,4,
\end{equation}
where $a=2R+d$, $\Psi _{l,m}^{0} (\vec{r})$ are wave functions of the ground $\Psi _{0,0}^{0} \equiv \Psi _{1s}^{0} $ or first excited electron states $\Psi_{1,m}^{0} =\{\Psi _{1p_{x} }^{0} ,\Psi _{1p_{y} }^{0} ,\Psi _{1p_{z} }^{0} \}\equiv \Psi _{1p}^{0} $ of an isolated quantum dot,
\[A(\vec{k})={\sqrt{a}}/{\sqrt{N}\left[1+\sum \limits_{\vec{n}}\re^{\ri\vec{k}\vec{n}}\int \Psi _{l,m}^{0*} (\vec{r}-\vec{n})\Psi_{l,m}^{0} (\vec{r})\rd\vec{r}\right]^{1/2}}\]
is the normalization constant. In case of $\nu =1$, we have an electron wave function of the 1$s$-like band, and in case of $\nu =2,3,4$ --- three 1$p$-like bands.
In the nearest-neighbor approximation, energy $E(\vec{k})$ can be found from the dispersion equation for electrons of the superlattice:
\[\sum _{\nu }C_{\nu } \left\{ \left[E_{\nu '}^{0} -E(\vec{k})\right]\left[\delta _{\nu \nu '}+2\sum _{i=1}^{3}A_{\nu \nu '}^{i} \cos (k_{i} a_{i} ) \right]+2\sum _{i=1}^{3} \left[ B_{\nu \nu '}^{i} +P_{\nu \nu '}^{i} \cos (k_{i} a_{i} )\right] \right\} =0,\]
where the following notations are introduced:
\begin{align}
A_{\nu \nu '}^{i} &=\int \phi _{\nu '}^{*} \left({\vec{r}}-{\vec{\emph{a}}}_{i} \right)\phi _{\nu } \left({\vec{r}}\right) \rd{\vec{r}},\nonumber\\
B_{\nu \nu '}^{i} &=\int \phi _{\nu '}^{*} \left({\vec{r}}\right)V\left(|{\vec{r}}-{\vec{\emph{a}}}_{i} |\right)\phi _{\nu }^{} \left({\vec{r}}\right) \rd{\vec{r}},\nonumber\\
P_{\nu \nu '}^{i} &=\int \phi _{\nu '}^{*} \left(\vec{r}-{\vec{\emph{a}}}_{i} \right)V\left(|\vec{r}-{\vec{\emph{a}}}_{i}|\right)\phi _{\nu }^{} \left( \vec{r}\right) \rd{\vec{r}}.\nonumber
\end{align}

To calculate the absorption coefficient according to equation~(\ref{alfa}), we substitute the summation over the wave vector with integration over $\vec{k}$ within the Brillouin zone. Using the properties of the $\delta$-function and taking into account the existence of two poles (for positive and negative values of the wave vector projections, they give the same contribution), after appropriate transformations, equation~(\ref{alfa}) is rewritten
\begin{equation}
\label{alfa1}
\alpha \left(\omega \right)=\frac{4\left(2\piup e\hbar \right)^{2} n_{0} }{a_{0}^{2}m_{0}^{2} \omega c\sqrt{\varepsilon } } \sum _{\nu =2,3,4}\left[f_{1} \left(\omega \right)-f_{\nu } \left(\omega \right)\right]a_{1}^{*} (\omega )a_{\nu }^{} (\omega )\eta _{}^{*} (\omega )\eta (\omega )/\left|\gamma (\omega )\right|.
\end{equation}

In equation~(\ref{alfa1}) we use the following notations:
\[a_{\left\{\substack{1s\\1p}\right\}} (\omega )=\left\{1+2\cos \left[ak_{0} \left(\omega \right)\right]\int \psi_{\left\{\substack{1s\\1p}\right\}}^{0*} \left(\vec{r}+\vec{m}_{0} \right)\psi_{\left\{\substack{1s\\1p}\right\}}^{0} \left(\vec{r}\right)\rd\vec{r} \right\}^{-1} ,\]
\begin{equation}
\label{eta}
\eta (\omega )=\int \psi _{1p}^{0*} \left(\vec{r}\right)\vec{\xi }\vec{\nabla }\psi _{1s}^{0} \left(\vec{r}\right)\rd\vec{r} +2\cos \left[ak_{0} \left(\omega \right)\right]\int \psi _{1p}^{0*} \left(\vec{r}+\vec{m}_{0} \right)\vec{\xi }\vec{\nabla }\psi _{1s}^{0} \left(\vec{r}\right)\rd\vec{r} ,
\end{equation}
$\gamma (\omega )=(E_{1p} -E_{1s} -\hbar \omega )' |_{k_{0} \left(\omega \right)}$,
$n_{0}$ is the concentration of QD chains $\left( n_{0}=N/S \right)$,  $k_{0} \left(\omega \right)$ is a positive pole of the $\delta$-function, $a_{0}$ is a lattice constant of GaAs. Equation~(\ref{eta})  contains a scalar product $\vec{\xi }\vec{\nabla }$ that needs to be specified. We consider several possible cases of polarization of the incident wave.

If the light wave is directed in parallel to the QD chains ($\vec{\chi }\parallel Oz$) [see figure~\ref{Figure 2.}~I(a)], we get $\vec{\xi }\vec{\nabla }=\frac{\partial }{\partial x} $ in case of the linearly polarized wave and $\vec{\xi }\vec{\nabla }=\frac{\partial }{\partial x} \pm \ri\frac{\partial }{\partial y} $ in case of the circularly polarized wave [see figure~\ref{Figure 2.}~II(a)]. If the light wave is directed perpendicular to the QD chain system ($\vec{\chi }\bot Oz$), then in case of a linearly polarized wave $\vec{\xi }\upuparrows Oz$ [see figure~\ref{Figure 2.}~I(b)] we get $\vec{\xi }\vec{\nabla }=\frac{\partial }{\partial z} $. The analogous geometry of the physical system in case of a circularly polarized wave [see figure~\ref{Figure 2.}~II(b)] gives $\vec{\xi }\vec{\nabla }=\frac{\partial }{\partial z} \pm \ri\frac{\partial }{\partial x} $.

\section{Analysis of numerical calculations}

Specific numerical calculations are performed for a GaAs/Al$_{x}$Ga$_{1-x}$As heterosystem. Its basic physical quantities depend on the aluminium concentration $x$ \cite{Boichuk15}. In the proposed model, we consider the dependence of the intersubband light absorption coefficient on the properties of the incident electromagnetic wave. We consider the system of one-dimensional QD superlattices. Their sizes permit the existence of exclusively the $s$- and $p$-like under-the-barrier electronic subbands. At the same time, spherical quantum dots, connected by tunnel effect and uniformly distributed in the same direction, are located at distances of the order of the GaAs crystal lattice parameter \cite{Boichuk15}. From equation~(\ref{alfa1}) it is seen that the absorption coefficient $\alpha (\omega )$ is determined by distribution functions of electrons in the $1s$-like and $1p$-like subbands, i.e., occupation density of these subbands. The integral parameter which defines the equilibrium state of the electron subsystem is the Fermi level. Its value is determined by concentration and type of doping donor impurities in the matrix of the heterosystem. First, we study the dependence of the function $\alpha =\alpha (\omega )$ on different values of $\mu $.

\begin{figure}[!t]
\begin{center}
  \includegraphics[width=0.7\linewidth]{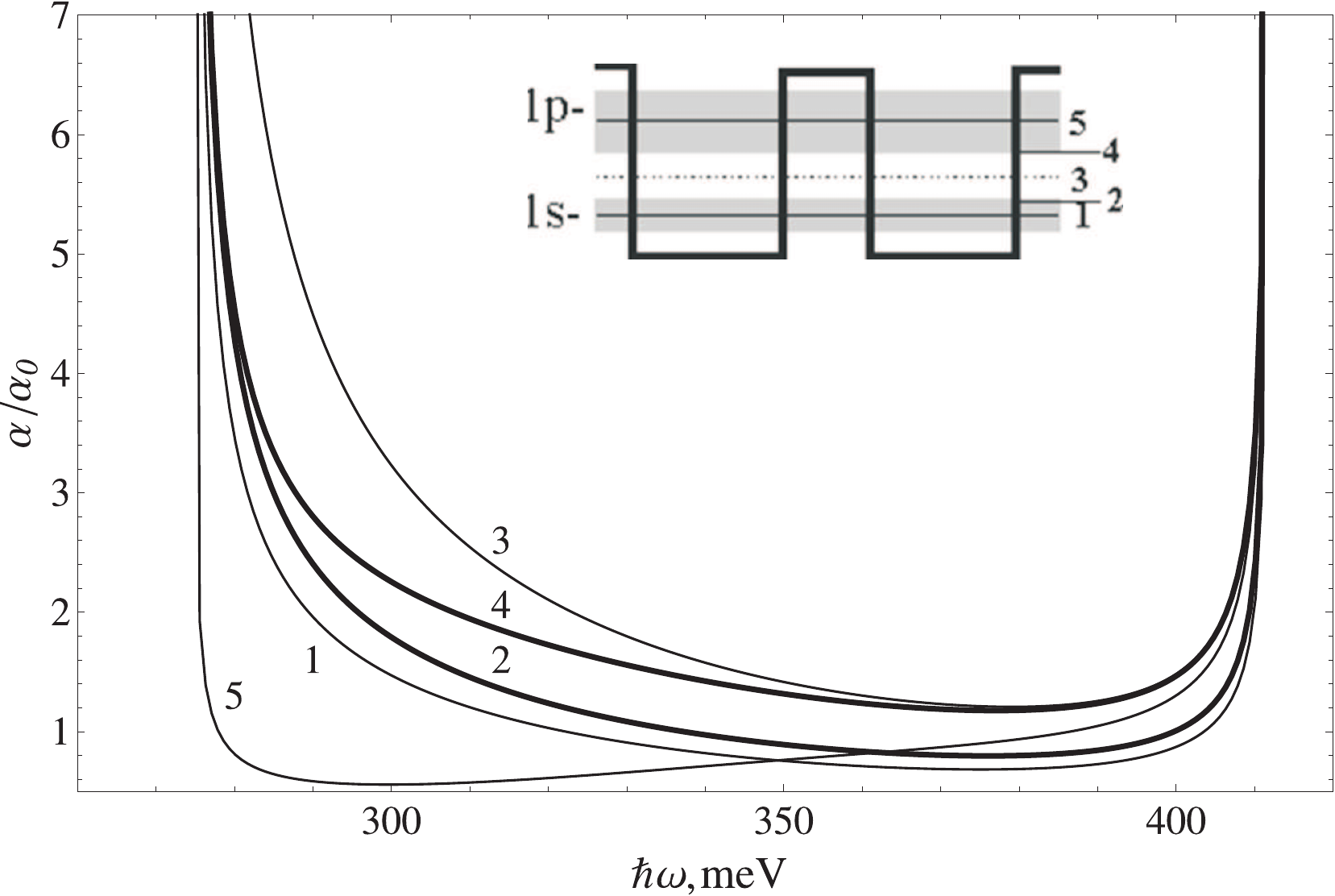}
  \caption{The relative coefficient of intersubband light absorption in case of linear polarized electromagnetic wave incident perpendicular to the QD chain at different values of the chemical potential: $-643.13$~meV --- curve~1, $-633.45$~meV --- curve~2, $-496.04$~meV --- curve~3, $-358.64$~meV --- curve~4, and $-299.89$~meV --- curve~5. The number of curves in figure corresponds to the number of Fermi level placing on the diagram.}\label{Figure 3.}
\end{center}
\end{figure}

Figure~\ref{Figure 3.} represents the relative coefficient $\alpha/\alpha_0$ of linearly polarized light incident perpendicularly to the spherical QD chain of radius $R=30$~{\AA} at $T = 300$~K, where $\alpha_0=4\left(2\piup e\hbar \right)^{2} n_{0} /[m_{0}^{2}a_{0}^{2} c\sqrt{\varepsilon} (E_{1p} -E_{1s})^2]$. The distance between quantum dots in the superlattice is $6$~{\AA}. Curve~1 corresponds to the Fermi level equal to the energy of the middle of the $s$-subband ($-643.13$~meV), curve~2 is the maximum of the $s$-subband ($-633.45$~meV), curve~3 is the average value between the $s$- and $p$-subbands ($-496.04$~meV), curve~4 is the minimum of the $p$-subband ($-358.64$~meV), and curve~5 is the energy of the middle of the $p$-subband ($-299.89$~meV). As is seen from the figure, curves~1--5 are of non-monotonous character within the absorption band and follow to infinity at its edges at energy values of $274.81$~meV and $411.66$~meV. The light absorption coefficient $\alpha =\alpha (\omega )$ rises with an increase of the Fermi energy $\mu $ from the $s$-subband to the average value between the $s$-, $p$-subbands region (curves~1--3). Curve~3, which is the highest, corresponds to the Fermi level equal to the average energy between the $s$- and $p$-subbands. However, changing the Fermi energy to the energy that corresponds to the $p$-subband (curves~$4,5$) causes a reduction of the coefficient. Thus, optimum conditions for light absorption in the heterosystem can be tuned by changing the concentration and type of the doping donor impurities (which change the Fermi level position) in the matrix.

\begin{figure}[!t]
\begin{center}
  \includegraphics[width=0.7\linewidth]{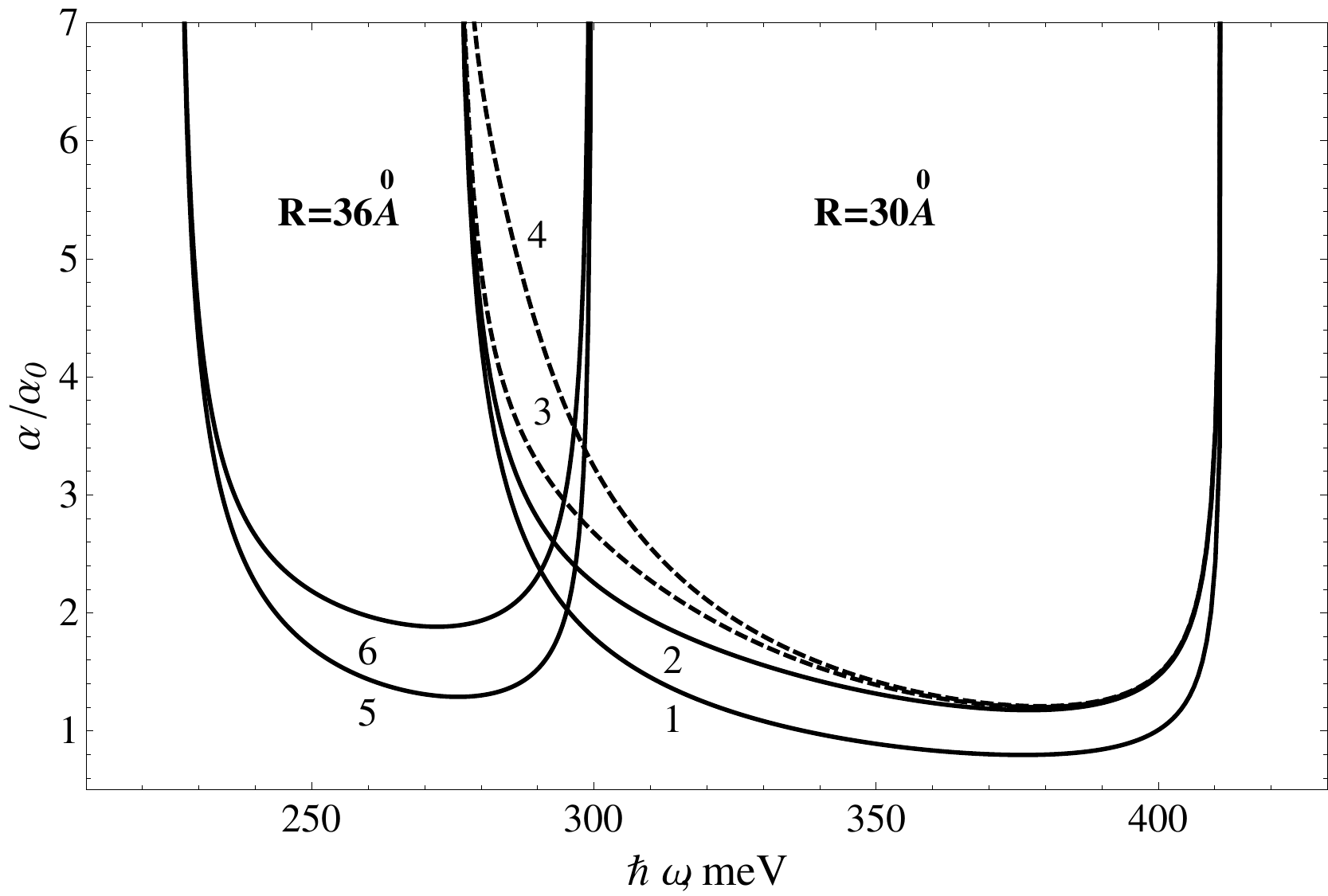}
  \caption{The absorption coefficient in case of linearly polarized light incident perpendicular to the chain at different QD radii: $R=30$~{\AA} --- curves~1--4,
   $R=36$~{\AA} --- curves~$5,6$ (distances between QD is $d=6$~{\AA}) and temperatures: $T=300$~K --- solid curves,  $T=40$~K --- dashed curves,
   when $\mu=E_{1s_{\text{max}}}$ (curves~$1,3,5$) and $\mu=E_{1p_{\text{min}}}$ (curves~$2,4,6$).} \label{Figure 5.}
\end{center}
\end{figure}

Figure~\ref{Figure 5.} illustrates the dependence of the relative coefficient $\alpha/\alpha_0$ on the energy of incident photons at temperatures $T=300$~K (solid curves) and $T=40$~K (dashed curves). When $\mu $ equals the energy of both the $s$-miniband maximum in $\vec{k}$-space (curves~$1,3,5$) and $p$-miniband minimum in $\vec{k}$-space (curves~$2,4,6$), we see that the absorption coefficient increases with a decreasing temperature for all bands. This is evident at the low-frequency edge of the absorption band in contrast to the opposite one. It is seen that under constant conditions (i.e., temperature, chemical potential) an increase in QD radius results in a reduction of the absorption bandwidth and leads to its red shift (curves~$5,6$).

\begin{figure}[!t]
\begin{center}
   \includegraphics[width=0.7\linewidth]{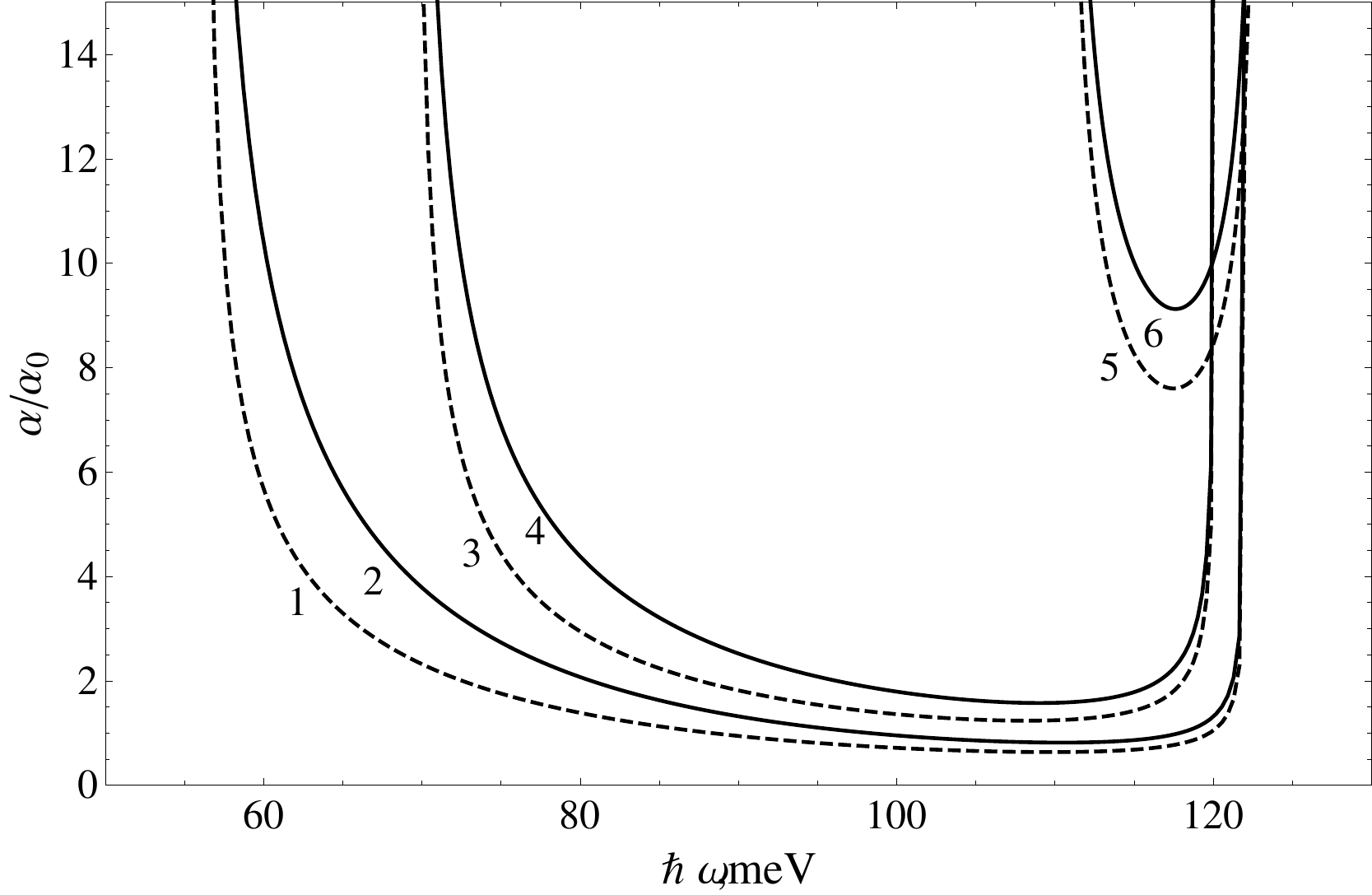}
  \caption{The relative absorption coefficient in case of linearly polarized light incident perpendicular to the GaAs/Al$_{x}$Ga$_{1-x}$As QD chain with $d=6$~{\AA} at different concentrations: curves~$1,2$ --- $x=0.3$, curves~$3,4$ --- $x=0.4$, curves~$5, 6$ --- $x=1.0$ and temperatures: $T=300$~K --- solid curves, $T=40$~K --- dashed curves; $\mu=\frac{1}{2}(\max{E_{1s}}+\min{E_{1p}})$.}\label{Figure 6.}
\end{center}
\end{figure}

The absorption bandwidth can also be changed by varying the composition of $x$ in the matrix Al$_{x}$Ga$_{1-x}$As where the GaAs QD array is embedded. In figure~\ref{Figure 6.}, the following notation is introduced: curves~$1, 2$ correspond to aluminium concentration $x=0.3$, curves~$3, 4$ to $x=0.4$, curves~$5, 6$ to $x=1.0$ and temperatures $T=300$~K (solid curve), $T=40$~K (dashed curve) at QD radius $R=60$~{\AA}. It is seen that increasing the concentration of aluminium $x$ leads to a decrease in the absorption bandwidth. This is due to a rise of the barrier height \cite{Boichuk15} and, consequently, due to a reduction of the widths of the $s$-, $p$-minibands. Thus, at aluminium concentration $x=0.3$, the widths of the $s$-, $p$-minibands are equal to 11.3~meV and 54.6~meV, respectively. A little change in its concentration to $x=0.4$ leads to a decrease of their values, 8.1~meV and 42.6~meV, respectively. Consequently, the function $\alpha =\alpha (\omega )$ within the absorption band increases.

\begin{figure}[!t]
\begin{center}
   \includegraphics[width=0.7\linewidth]{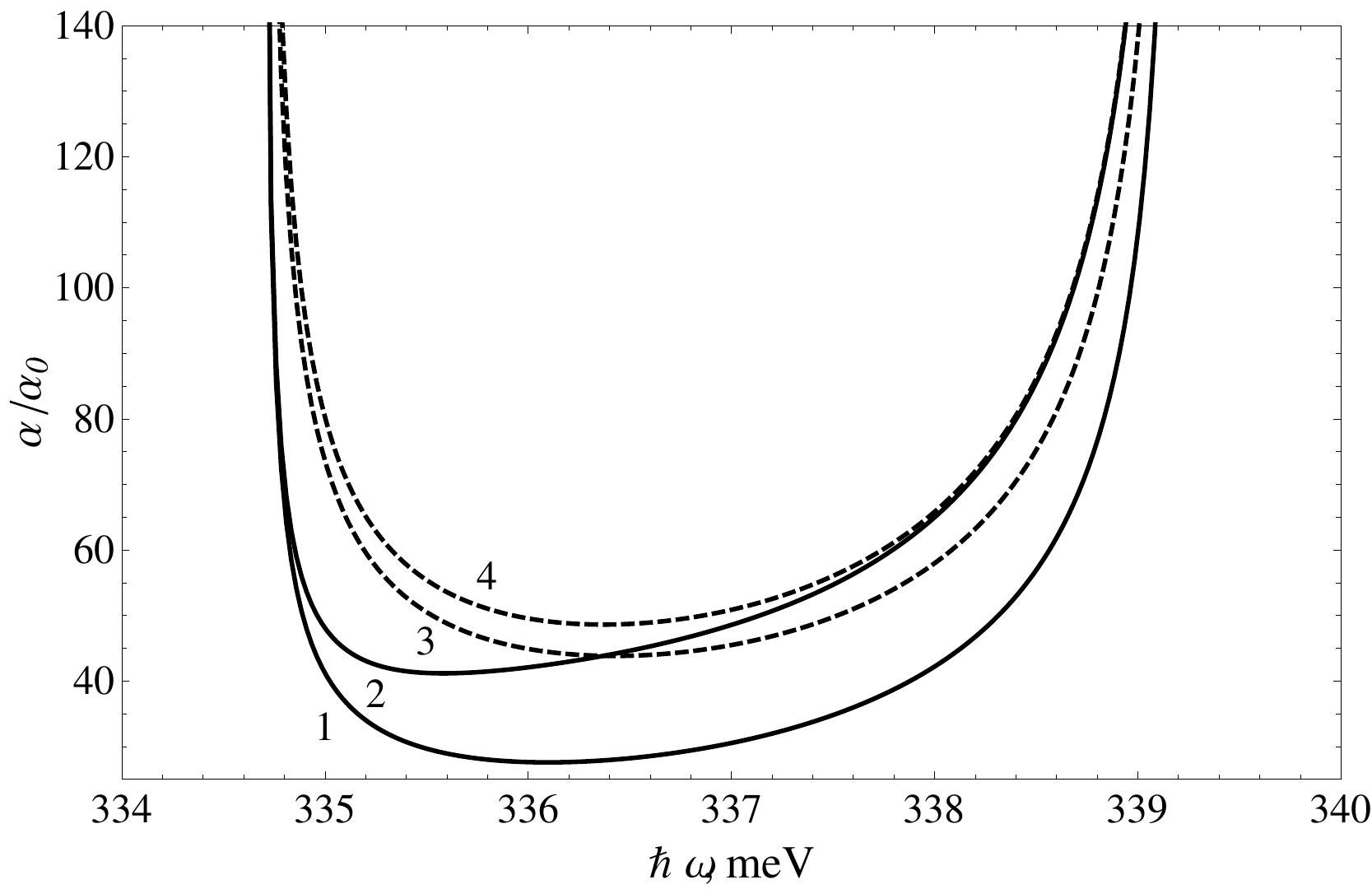}
  \caption{The relative absorption coefficient in case of linearly polarized light incident parallel to the QD chain at different temperatures: curves~$1, 3$ --- $T=300$~K, curves~$2, 4$ --- $T=40$~K.
  Radius GaAs QD $R=30$~{\AA} and distances between $d=6$~{\AA} in AlAs matrix.}
  \label{Figure 7.}
  \end{center}
\end{figure}

Figure~\ref{Figure 7.} displays the relative absorption coefficient of linearly polarized light for $\vec{\chi }\upuparrows Oz$, QD radius $R=30$~{\AA} at different temperatures. Curves~$1, 3$ correspond to room temperature and curves~$2, 4$ to $T=40$~K. In case of the chemical potential equal to the maximum value of the $s$-miniband, the dependence is shown by a solid curve; a dashed curve stands for the minimum value of the $p$-miniband. As is seen from the figure, all curves~1--4 display typical minima closer to the left-hand edge of the absorption band and reach maximum values at the edges of the absorption band. Calculations show that the width of the absorption band is much narrower than in the case of the light incident perpendicularly to the QD chain. This can be explained by the fact that in this case the electric field vector is perpendicular to the QD chain. Therefore, in case of the dipole approximation, the electron transition is allowed from the 1$s$-subband in the 1$p_{x}$-subband, which is much narrower than the 1$p_{z}$-subband (see figure~\ref{Figure 3.}).

\begin{figure}[!b]
\begin{center}
   \includegraphics[width=0.49\textwidth]{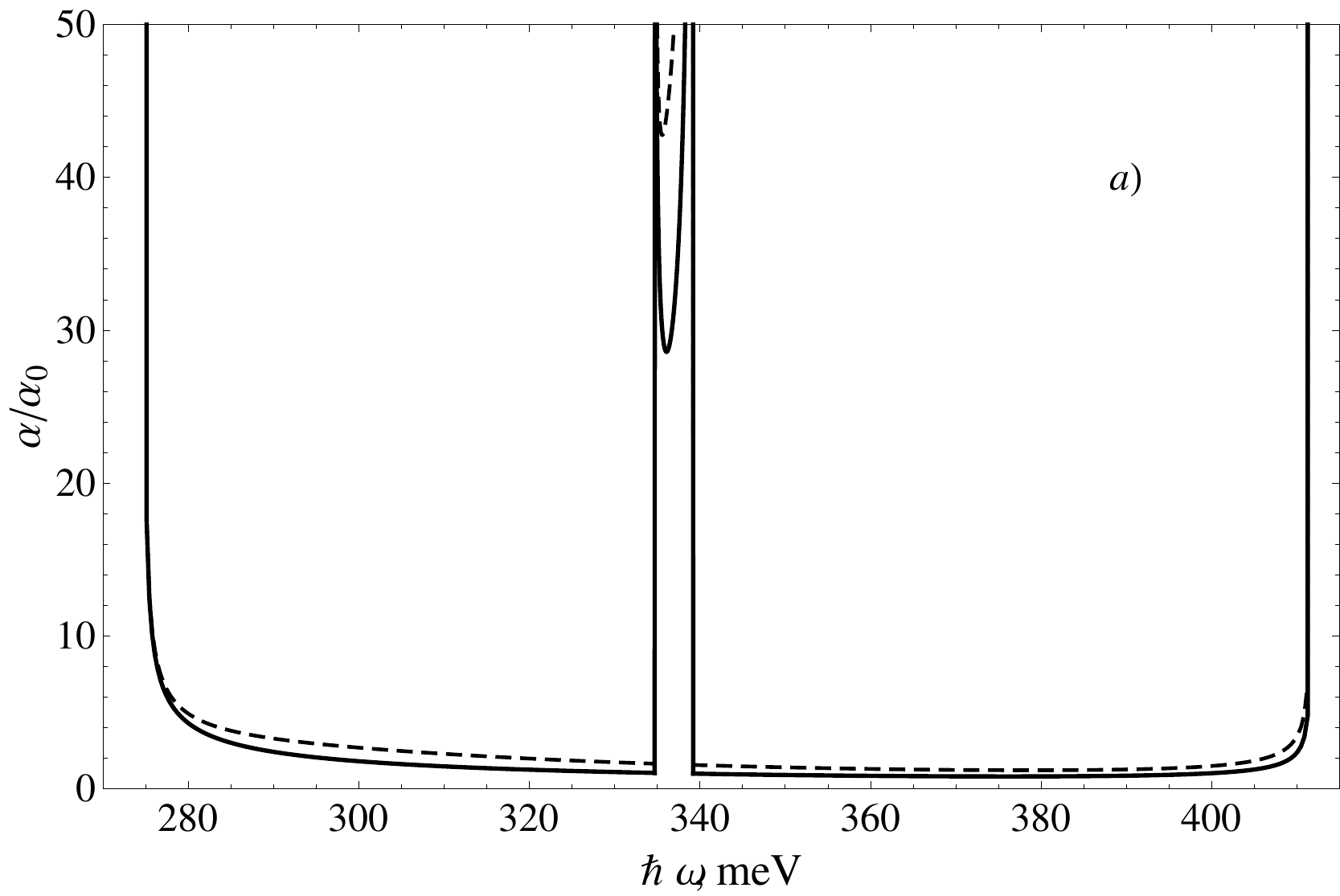}
   \includegraphics[width=0.49\textwidth]{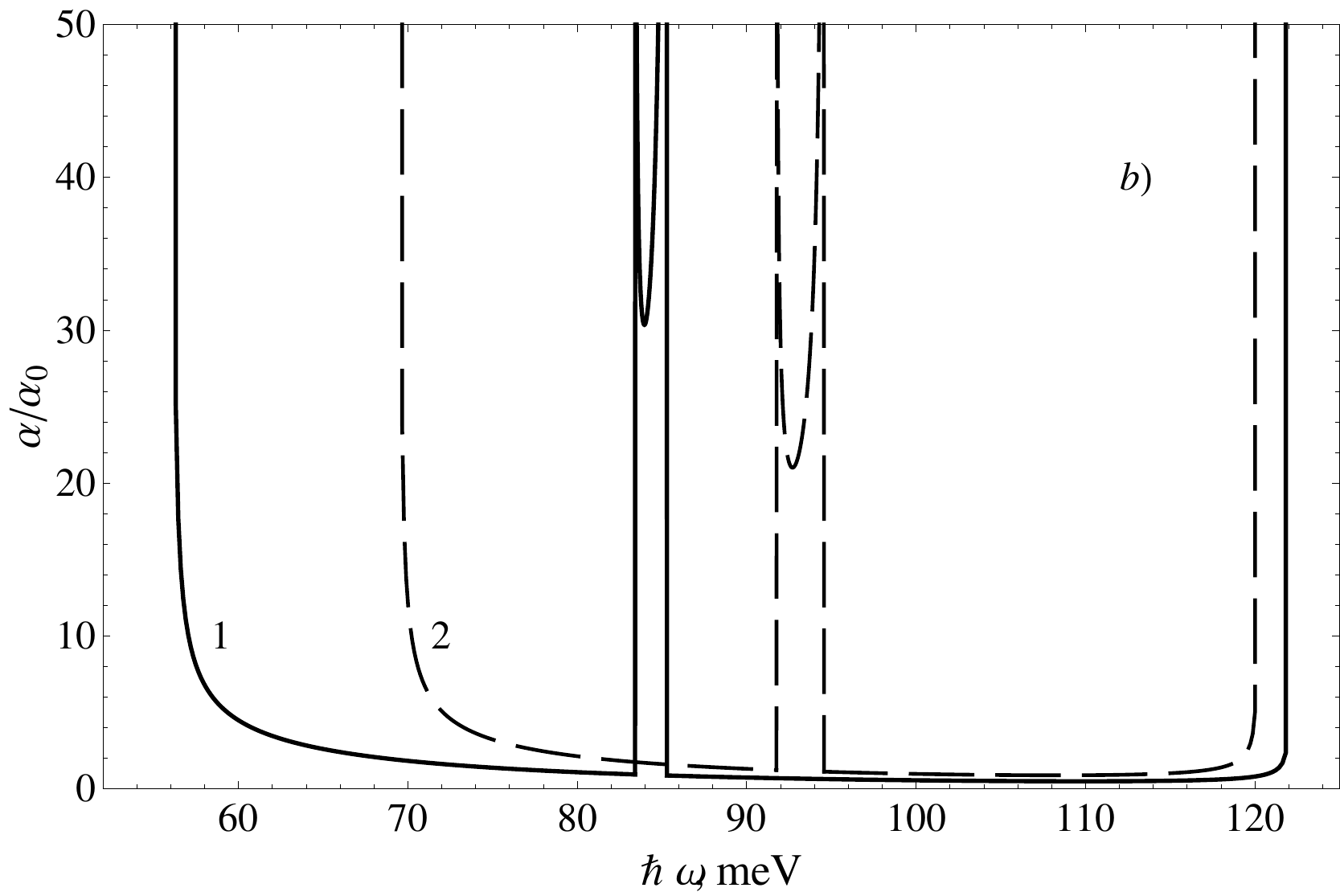}
   \caption{The absorption coefficient in case of circularly polarized light incident perpendicular to the GaAs/Al$_{x}$Ga$_{1-x}$As QD chain with $d=6$~{\AA}:
   (a) aluminium concentration in matrix $x=1$,  curve~1 --- $T=300$~K, curve~2 --- $T=40$~K, radius QD $R=30$~{\AA};
   (b) curve~1 --- $x=0.3$, curve~2 --- $x=0.4$; $T=300$~K,   $R=60$~{\AA}.} \label{Figure 8.}
\end{center}
\end{figure}

The paper also presents the case of illuminating the QD array by circularly polarized light (see figure~\ref{Figure 2.}~II). Figure~\ref{Figure 8.} shows the relative light absorption coefficient in case of circularly polarized light incidence perpendicular to the GaAs/AlAs QD chain ($R=30$~{\AA}) at different temperatures: curve~1 corresponds to room temperature, and curve~2 to $T=40$~K. The coefficient $\alpha $ reaches maximal values at the edges of the absorption band. Within the band, the function $\alpha =\alpha (\omega )$ grows with a decreasing temperature as is in the case of linear polarization of light (see figure~\ref{Figure 3.}). However, circular polarization involves an additional quasi-resonant band inside the absorption region. This is due to not only electronic transitions between the 1$s$- and 1$p_{z}$-subbands but also between the 1$s$- and 1$p_{x}$-, 1$p_{y}$-subbands. The change of aluminium concentration in the matrix, as is shown in figure~\ref{Figure 8.}~(b) for the GaAs/Al$_{x}$Ga$_{1-x}$As QD (a solid curve corresponds to $R=60$~{\AA}, $x=0.3$; a dashed curve stands for $x=0.4$), in its turn, the absorption bandwidth changes, similar to the case of linear polarization (see figure~\ref{Figure 6.}).

Calculations show that in case of electromagnetic wave incident parallel to the QD chain, the absorption coefficient does not depend on the light polarization method whether it is linear or circular. At the same time, changing the incidence direction to perpendicular leads to the appearance of a quasi-resonant band inside the absorption region. Its width and frequency correspond to the case of linearly polarized light directed in parallel to the spherical QD chain.

\section{Conclusions}

We considered intersubband electronic transitions in an array of non-interacting one-dimensional chains of the spherical semiconductor quantum dot system GaAs/Al$_{x}$Ga$_{1-x}$As. The coefficient of light absorption caused by transitions was calculated depending on the frequency and polarization of the incident light, Fermi level position, temperature, and concentration of aluminium in the matrix. It has been established that
\begin{itemize}
\item there are peaks of the light absorption coefficient at the edges of the absorption region;

\item  the relative absorption coefficient depends on Fermi level position and reaches its maximum at the center of the region between the $s$-, $p$-like subbands and slightly varies with temperature;

\item  in case of linear polarization of the light wave incident perpendicular to the chains, the width of the absorption band sharply increases (several times) in comparison with linearly polarized light oriented parallel;

\item  the width of the absorption band also increases with a decreasing quantum dot radius followed by a simultaneous blue shift;

\item  broadening of the absorption band with a decreasing concentration of aluminium (simultaneously with an increasing concentration of gallium) in the Al$_{x}$Ga$_{1-x}$As matrix is caused by reducing the barrier height and, thus, increasing the miniband widths;

\item  the absorption coefficient of light incident parallel to the quantum dot chain does not depend on the method of polarization (linear or circular), while the change in the direction of polarization to perpendicular leads to the appearance of a quasi-resonant band in the absorption region.
\end{itemize}

\ukrainianpart

\title{Коефіцієнт поглинання світла масивом впорядкованих ланцюжків сферичних квантових точок }
\author{В.І. Бойчук, І.В. Білинський, Р.І. Пазюк}
\address{Кафедра теоретичної фізики і прикладної фізики та комп’ютерного моделювання, Дрогобицький державний педагогічний університет імені Івана Франка,
вул. Івана Франка, 24, 82100 Дрогобич, Україна}

\makeukrtitle

\begin{abstract}
Розглянуто міжпідзонні оптичні переходи в масиві невзаємодіючих одновимірних ланцюжків сферичних квантових точок напівпровідникової системи GaAs/Al$_{x}$Ga$_{1-x}$As. Обчислено коефіцієнт поглинання, зумовлений цими переходами, в залежності від частоти та поляризації падаючого світла, залягання рівня Фермі та температури. Встановлено наявність максимумів коефіцієнта поглинання на краях області поглинання в результаті міжпідрівневих переходів за встановленими правилами відбору та його залежність від положення рівня Фермі. Показано, що максимальне значення $\alpha$ досягається в центрі області між $s$-, $p$-подібними зонами і слабо змінюється з температурою. При зміні напрямку падіння лінійно поляризованої світлової хвилі з перпендикулярного 1D-НСКТ на паралельний ширина смуги поглинання різко зменшується. Ширина смуги поглинання також зростає і при зменшенні радіуса КТ з одночасним зсувом її в довгохвильову область. Отримано, що коефіцієнт поглинання не залежить від способу поляризації (лінійно чи циркулярно) світла, падаючого паралельно ланцюжку КТ, тоді як зміна напрямку поляризації на перпендикулярний призводить до появи квазірезонансної смуги всередині області поглинання. Проаналізовано залежність відносного коефіцієнта поглинання від концентрації Al в матриці GaAs/Al$_{x}$Ga$_{1-x}$As.

\keywords квантова точка, надґратка, електронні стани, густина станів, коефіцієнт поглинання

\end{abstract}

\end{document}